\documentclass[twocolumn,secnumarabic,amssymb,nobibnotes,aps,pra]{revtex4-1}
\usepackage{amssymb,amsmath}
\usepackage{enumerate}
\usepackage{graphicx}
\newcommand{\bra}[1]{\langle #1|}
\newcommand{\ket}[1]{|#1\rangle}
\newcommand{\braket}[2]{\langle #1|#2\rangle}
\newcommand{\ketbra}[2]{|#1 \rangle \langle #2|}

\def\tr{\,{\rm Tr}}

\def\i{\,{\rm i}}

\begin{document}
\title{Purification-based metric to measure the distance between quantum states and processes}
\author{Trist\'an M. Os\'an and Pedro W. Lamberti}        
\affiliation{Facultad de Matem\'atica, Astronom\'{\i}a y F\'{\i}sica, Universidad Nacional de C\'ordoba and CONICET, Medina Allende s/n, Ciudad Universitaria, X5000HUA C\'ordoba, Argentina.}

\date{\today}          


\begin{abstract}
In this work we study the properties of an purification-based entropic metric for measuring the distance between both quantum states and quantum processes. This metric is defined as the square root of the entropy of the average of two purifications of mixed quantum states which maximize the overlap between the purified states. We analyze this metric and show that it satisfies many appealing properties, which suggest this metric is an interesting proposal for theoretical and experimental applications of quantum information.
\end{abstract}

\maketitle 

\section{Introduction}

Quantum Information Processing is intended to develop new forms and procedures of computation and cryptography beyond the possibilities of classical devices. Thus, a significative quantity of new algorithms, communications protocols, and suggestions for physical implementations of theoretical concepts has been proposed \cite{Pitt,Bouw,Kempe,Deutsch85,Deutsch92,Shor94,Grover,Ekert,Shor97,Cleve,Childs,Niel}. As a consequence, quantum information continues to be a topic of major interest for current research.\\
Most important noiseless quantum communication protocols such as teleportation, super-dense coding, including their coherent versions, and entanglement distribution rely on the assumption that noiseless resources are available. For example, the entanglement distribution protocol assumes that a noiseless qubit channel is available to generate a noiseless entangled bit (ebit). This idealization allows to develop the main principles of the protocols without the need to take into account more complicated issues. However, in practice, quantum protocols do not work as expected in the presence of noise.\\
In order to protect quantum information from noise some strategies have been proposed, like quantum error-correcting codes and fault-tolerant quantum computation \cite{Niel,Shor95,Steane,Gott97}. In this regard, a large number of error-correcting codes have been developed. For example, a promising approach is to use topological error-correcting codes to store quantum information safely by associating it with some topological property of the system \cite{Presk,Kitaev,Bombin06}. This strategy works in such a way to make quantum information resilient against the effects of noise. A recent proposal in this area can be found in references \cite{Bombin12,Gott12}. Another particularly fruitful strategy seems to be the group-theoretical structure known as ``stabilizer codes'' \cite{Gott97}.\\
Despite the existence of these strategies to protect quantum information from noise, in many practical cases it is desirable to have the means to quantify how much a quantum system is effectively affected by a disturbance, no matter how small. In other words, it is important to have a procedure to determine how close to expected a real quantum system is working.
The simplest way to do so is to compare the output state of the quantum system, thought as ideal, with the output state of the real system using a distance measure between them. For example, suppose that a quantum information processing protocol should ideally produce some quantum state represented by a density operator $\rho$, but the actual output of the protocol is a mixed quantum state represented by a density operator $\sigma$, then, a distance measure $D(\rho,\sigma)$ should be provided to indicate how close the ideal output of a quantum process is to the actual output.\\
One of the most important features of quantum mechanics is that, in general, two arbitrary quantum states cannot be determined with certainty. For example, if two pure states are non-orthogonal they cannot be perfectly distinguished. Only orthogonal states can be discriminated unambiguously. Therefore, in order to provide a way to determine how well a quantum protocol is working, distance measures need to be devised to allow us to determine how close two quantum states or two quantum processes are to each other.\\
A variety of distance measures have been developed for this purposes, like {\em trace distance}, {\em Fidelity}, {\em Bures distance}, {\em Hilbert-Schmidt distance}, {\em Hellinger distance} and {\em Quantum Jensen-Shannon divergence}, just to name a few \cite{Niel,Bengt,Uhlm,Jozsa,Bures,Dodo,Luo,Majtey05}.\\
Quantum processes can be represented by means of positive and trace-preserving maps $\mathcal{E}$ defined on the set of density operators belonging to $\mathcal{B(\mathcal{H})}_1^+$, that is, the set of positive trace one operators $\rho$ on a Hilbert space $\mathcal{H}$.\\
We say that the map $\mathcal{E}$ is {\em monotonous under quantum operations} with respect to a given distance $D(\rho,\sigma)$, or {\em contractive} for short, if

\begin{equation}
D(\mathcal{E}(\rho),\mathcal{E}(\sigma)) \leq D(\rho,\sigma)
\end{equation}

In particular, when $\mathcal{E} = \mathcal{E}_t$ is a completely positive quantum dynamical semigroup such that $\rho(t) = \mathcal{E}_t\rho(0)$, then contractivity
means that 

\begin{equation}
D(\mathcal{E}(\rho(t)),\mathcal{E}(\sigma(t)) \leq D(\rho(t'),\sigma(t')),\:\:\:\: {\rm for}\:\: t > t'
\end{equation}

The physical meaning of last equation is that the distance between two quantum processes cannot increase in time and the distinguishability of any pair of states cannot increase beyond an initial value.\\
A case of particular interest is a quantum open system \cite{Davies,Breuer}. A real quantum system $Q$, like a system intended to perform a quantum information processing task, is always in interaction with its environment $E$. This interaction inevitably has an influence on the state of the quantum system, causing losses on the information encoded in the system. The quantum system $Q$ can not be treated as a closed system anymore when interactions with the outside world are occurring. This kind of systems is known as {\em open quantum system}. Time evolution of the open system $Q$ cannot in general be described by an unitary operator acting on the Hilbert space $\mathcal{H}_Q$ of $Q$. As the total system is assumed to be closed, it will evolve with a unitary operator $U(t)$ acting on the total Hilbert space $\mathcal{H}_{QE} = \mathcal{H}_{Q}\otimes \mathcal{H}_{E}$. In many cases, as we are interested in extracting information on the state of the system $Q$ at some later time $t > 0$, we perform a partial trace over the environment $E$ to obtain the reduced state of the system $Q$ alone,

\begin{equation}
\rho_Q(t) = \tr_{E} \left[ U(t) \rho_{QE} U^\dag(t) \right]
\end{equation}

Given two initial states $\rho_{QE}$ and $\sigma_{QE}$ of the composite system, the distance between the corresponding reduced states $\rho_{Q}$ and $\sigma_{Q}$ at a given time $t>0$ can be contractive with respect to some distance measures but not necessarily to all of them. For instance, when an open quantum system $Q$ and its environment $E$ are initially prepared in an uncorrelated state, the reduced dynamics is completely positive and contractive, therefore, the distance $D(\rho_{Q},\sigma_{Q})$ between two states can approximate to zero when the open system is reaching a unique steady state (This is the case, for example, when dynamics is of the relaxing type). However, contractivity of quantum evolution can show a breakdown when system and environment are initially correlated. Effects induced by such correlations have been studied in different contexts \cite{Pech94,Pech95,Stelm01,Boul04,Jord04,Rom04,Smir10,Tan11,Ban11,Moroz}. As a result, contractivity turns out to be not an universal feature but rather depends on the correlations between the system and its environment and also, in general, on the particular choice of the distance measure.\\
Experiments on initial system-environment correlations can be found in Refs. \cite{Li11} and \cite{Smir11}. Examples of an exact reduced dynamics which fail contractivity with respect to the trace distance are presented in Refs. \cite{Dajka10} and \cite{Laine10}.
An increase of the distance between the states of the reduced system $Q$ can be interpreted in terms of an exchange of information between the system $Q$ and its environment $E$. For example, an increment of the distance above its initial value can be interpreted as information locally inaccessible for the system $Q$ at the beginning which was transferred to it later. As a result, this flow of information increases the distinguishability between reduced system states. Possibly, this process could be used to devise experimental schemes for detection of initial correlations between an open quantum system and its environment.\\
Dajka, {\L}uczca and H\"anggi \cite{Dajka11} performed a comparative study of different distances measures between quantum states in the presence of initial qubit-environment correlations. In that work they show that the correlation-induced distinguishability growth is not generic with respect to distance measures, but distinctly depends on the particular choice of the distance measure. Their results indicate that an increase of a distance measure above its initial value constitutes no universal property. Dynamics behavior upon evolving time strongly depends on the employed distance measure.\\
To the present, there is not a unique or ideal measure of distinguishability between quantum states or quantum process. Moreover, different distance measures can be useful depending on the particular application, whether a theoretical one, like a bound of what can be physically feasible for a given process, or the measurement of a quantum protocol experimentally implemented.\\
In a previous work \cite{Lamb2009}, a metric $D_E$ based on the physical concepts of entropy and purification of a mixed state was introduced \cite{noteDN}. Some useful properties of $D_E$ were studied and, in addition, it was demonstrated that $D_E$ is a true metric between quantum states.\\
In this work we extend the study of the properties of $D_E$ and we also derive an alternative fidelity measure $F_E$ for the degree of similarity between quantum states. We investigate the properties of $F_E$ and show that it shares the main properties of Uhlmann-Jozsa fidelity $F$ \cite{Uhlm,Jozsa}. In addition, as a main result, we derive from $D_E$ a distance measure $\Delta_E$ between quantum processes which turns to have many interesting properties for applications in quantum information.\\
This paper is organized as follows. In Sec. \ref{DMQIP}, we briefly introduce the criteria that should be satisfied for a suitable metric between quantum processes. In Sec. \ref{DescQP} we outline two approaches to describe quantum processes, operator-sum representation and Jamio{\l}kowski isomorphism. These descriptions will allow us later to derive from $D_E$ a distance measure $\Delta_E$ between quantum processes and to study its properties. 
In Sec. \ref{DEMETRIC} we describe the distance $D_E$ and we study its properties. In Sec. \ref{ALTFE} we introduce the alternative fidelity measure $F_E$. Next, in Sec. \ref{DeltaE}, we show how a measure of distance between quantum processes can be derived from $D_E$ and we study its properties. Finally, we summarize our main results in Sec. \ref{CONCREM}. In the appendix, with the purpose of making this work self-contained, we survey some important properties of the Uhlmann-Jozsa fidelity $F$ that will be used in order to prove some properties of $D_E$.

\section{\label{DMQIP}Distance measures in quantum information processing}

As stated before, there is not a unified criterium to chose a measure of distance between quantum states and quantum processes. However, some guidelines can be provided based on physical grounds. In this work we have chosen ourselves to follow the work of Gilchrist, Langford and Nielsen \cite{Gilch} as a guideline of what criteria a good measure of distance between quantum processes should satisfy.\\
Suppose $\Delta$ is a good measure of the distance between two quantum processes. Such processes are described by maps between input and output quantum states, e.g., $\rho_{out}= \mathcal{E}(\rho_{in})$, where the map $\mathcal{E}$ is a completely positive trace-preserving map (CPTP-map) also known as a {\em quantum operation} \cite{noteCPm1}. Physically, $\Delta(\mathcal{E},\mathcal{F})$ may be thought of as a measure of error in quantum information processing when it is desired to perform an ideal process $\mathcal{E}$ and the actual process $\mathcal{F}$ is obtained instead. In addition, $\Delta(\mathcal{E},\mathcal{F})$ can be interpreted as a measure of distinguishability between the processes $\mathcal{E}$ and $\mathcal{F}$. 

Bearing in mind the work of Gilchrist, Langford and Nielsen \cite{Gilch}, we will look for a measure $\Delta$ between quantum processes which should satisfy the following criteria, motivated by both physical and mathematical matters \cite{Notecriteria}:

\begin{enumerate}
\item \label{p1}{\em Metric}: $\Delta$ should be a metric, i.e., for any quantum processes $\mathcal{E}$, $\mathcal{F}$ and $\mathcal{G}$ the following properties should be satisfied: 
\begin{enumerate}[(i)]
\item {\em Non-negativity}, $\Delta(\mathcal{E},\mathcal{F}) \geq 0$ with $\Delta(\mathcal{E},\mathcal{F}) = 0$ if and only if $\mathcal{E} = \mathcal{F}$
\item {\em Symmetry}, $\Delta(\mathcal{E},\mathcal{F}) = \Delta(\mathcal{F},\mathcal{E})$
\item {\em Triangle inequality} $\Delta(\mathcal{E},\mathcal{F})\leq \Delta(\mathcal{E},\mathcal{G}) + \Delta(\mathcal{G},\mathcal{F})$.
\end{enumerate}
\item \label{p4} {\em Physical interpretation}: $\Delta$ should have a well-motivated physical interpretation.
\item \label{p5} {\em Stability} \cite{Aharon}: $\Delta(\mathcal{I} \otimes \mathcal{E}, \mathcal{I} \otimes \mathcal{F}) = \Delta(\mathcal{E},\mathcal{F})$ where $\mathcal{I}$ represents the identity operation on an extra Hilbert of arbitrary dimension. This ancillary Hilbert space could be associated to a quantum system or to a convenient mathematical construct. The physical meaning behind this property is that unrelated ancillary quantum systems do not change the value of $\Delta$.
\item \label{p6} {\em Chaining}: $\Delta(\mathcal{E}_2 \circ \mathcal{E}_1, \mathcal{F}_2 \circ \mathcal{F}_1) \leq  \Delta(\mathcal{E}_1,\mathcal{F}_1)+\Delta(\mathcal{E}_2,\mathcal{F}_2)$. This property just means that for a process composed of several steps, the total error is bound by the sum of the errors originated in the individual steps.
\end{enumerate}

From a mathematical viewpoint, it is evident that a character of true Metric is a basic requirement for a suitable distance measure. Besides, the metric character of a distance could be considered as essential to check on the convergence of iterative algorithms in quantum processing \cite{Galindo}. In addition, chaining and stability criteria are key properties to estimate the error in complex tasks of quantum information processing which can be split into sequences of simpler component operations. In this case, a bound on the total error can be found by analyzing each single step of a process.


\section{\label{DescQP}Describing Quantum Processes}

\subsection{\label{OPSR}Operator-sum representation}

Quantum operations describe the most general physical processes that may occur in a quantum system \cite{Niel,Bengt,Jaeger}, including unitary evolution, measurement, noise, and decoherence. Any quantum operation can be expressed by means of an operator-sum representation relating an input state $\rho$ with the output state $\mathcal{E}(\rho)$ in the form \cite{Kraus71,Kraus83,Niel,Bengt,Jaeger}

\begin{equation}
\label{osr}
\mathcal{E}(\rho) = \sum_j K_j \rho K_j^\dag
\end{equation}

where the operators $K_j$ are known as Kraus operators or {\em operation elements}, and satisfy the condition $\sum_j K_j^\dag K_j \leq I$. Particularly, when Kraus operators satisfy the equation

\begin{equation}
\label{TP}
\sum_j K_j^\dag K_j = I
\end{equation}

the process $\mathcal{E}(\rho)$ is a completely positive trace-preserving map (CPTP-map) and maps density matrices into density matrices. Physically, this corresponds to the requirement that $\mathcal{E}$ represents a physical process without post-selection \cite{postsel}. An important remark is that the operation elements $\lbrace K_j \rbrace$ completely describe the effect of the quantum process on the input state $\rho$.\\
Relation \ref{TP} is a completeness relation because $K_j$ and $K_j^\dag$ do not necessarily commute.
If additionally, the operation elements $K_j$ satisfy

\begin{equation}
\sum_j K_j K_j^\dag = I
\end{equation}

then, the CPTP-map is said to be a {\em unital} map, this means, a map for which $\mathcal{E}(I) = I$. One example of such a map is the qubit-depolarizing channel whereas a negative example is provided by the amplitude-damping channel \cite{Niel,Bengt,Jaeger}. If the operator decomposition of a CP-map satisfies both these conditions the map is doubly stochastic. The operator decomposition of a quantum operation is not unique. In particular, any two sets of operators $K_j$ related to each other by unitary transformations equally well represent the same operation $\mathcal{E}(\rho)$.

\subsection{\label{JAMISO}The Jamio{\l}kowski isomorphism}

Jamio{\l}kowski isomorphism relates a quantum operation $\mathcal{E}$ to a quantum state, $\rho_{\mathcal{E}}$, by the following equation \cite{Jamiol,Bengt,Vedral}

\begin{equation}
\label{jamiso}
\rho_{\mathcal{E}} = \left[ \mathcal{I} \otimes \mathcal{E} \right]\rho_\Psi
\end{equation}

where  $\rho_\Psi = \ketbra{\Psi}{\Psi}$ and 

\begin{equation}
\label{maxentang}
\ket{\Psi} = \frac{1}{\sqrt{d}}\sum_j \ket{j} \otimes \ket{j}
\end{equation}

is a maximally entangled state of the ({\it d}-dimensional) system with another copy of itself, and $\lbrace \ket{j} \rbrace$ is some orthonormal basis set.
Jamio{\l}kowski isomorphism works bidirectionally, i.e., the map $\mathcal{E}\rightarrow \rho_{\mathcal{E}}$ is invertible. Therefore,
the knowledge of $\rho_{\mathcal{E}}$ is equivalent to knowledge of $\mathcal{E}$ \cite{notejam}. As a consequence, this isomorphism allows to treat quantum operations using the same tools usually used to treat quantum states.

\section{\label{DEMETRIC}Purification-based entropic metric $D_E$}

Given two pure quantum states $\ket{\psi}$ and $\ket{\varphi}$, the distance $D_E(\ket{\psi},\ket{\varphi})$ introduced in Ref. \cite{Lamb2009}, is defined as \cite{noteDN}: 

\begin{equation}
\label{defDEpure}
D_E(\ket{\psi},\ket{\varphi}) \equiv \sqrt{H_N\left( \frac{\ket{\psi}\bra{\psi} + \ket{\varphi}\bra{{\varphi}}}{2} \right)}
\end{equation}

where $H_N(\rho)$ represents the von Neumann entropy given by:

\begin{equation}
H_N(\rho) = -\tr \left( \rho \log_2 (\rho) \right) = - \sum_i \lambda_i \log_2(\lambda_i)
\end{equation}

with $\lbrace \lambda_i \rbrace$ being the set of eigenvalues of the density operator $\rho$.\\

The distance $D_E$ emerges from the quantum Jensen-Shannon divergence $D_{JS}$ defined as \cite{Majtey05}:

\begin{equation}
\label{QJSD}
D_{JS}(\rho,\sigma) = H_N \left(\frac{\rho+\sigma}{2}\right) - \frac{1}{2} H_N(\rho) - \frac{1}{2}H_N(\sigma)
\end{equation}

Indeed, due to von Neumann entropy vanishes when evaluated in pure states $\rho = \ketbra{\psi}{\psi}$ and $\sigma = \ketbra{\varphi}{\varphi}$, the $D_{JS}$ reduces to

\begin{equation}
\label{QJSDpure}
D_{JS}(\ketbra{\psi}{\psi},\ketbra{\varphi}{\varphi}) = H_N \left(\frac{\ketbra{\psi}{\psi} + \ketbra{\varphi}{\varphi}}{2}\right)
\end{equation}

As a consequence, the distance $D_E$ verifies the identity

\begin{equation}
\label{DEvsQJSD}
D_E^2(\ketbra{\psi}{\psi},\ketbra{\varphi}{\varphi}) = D_{JS}(\ketbra{\psi}{\psi},\ketbra{\varphi}{\varphi})
\end{equation}

\begin{figure}[h!]
\includegraphics[scale=0.25]{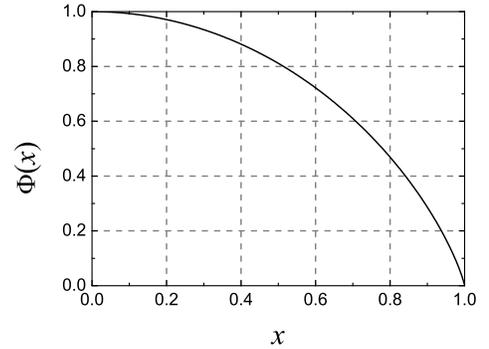}
\caption{\label{fig1}Plot of $\Phi(x)$ given by Eq. \ref{Phix}}
\end{figure}

After some algebra, it is possible to write $D_E$ in the form \cite{Lamb2009}

\begin{equation}
\label{DEPhix}
D_E(\rho,\sigma) = \sqrt{\Phi\left(|\braket{\psi}{\varphi}|\right)}
\end{equation}

where

\begin{equation}
\label{Phix}
\Phi(x) \equiv - \left( \frac{1 - x}{2}\right) \log_2 \left( \frac{1 - x}{2}\right) - \left( \frac{1 + x}{2}\right) \log_2 \left( \frac{1 + x}{2}\right)
\end{equation}

with $\Phi(x)$ being the Shannon entropy of a probability vector of size 2 and $x=|\braket{\psi}{\varphi}|$. From equation \ref{Phix} it is easy to see that $\Phi(x)$ is a {\em bounded} and {\em monotonic decreasing} function of $x$ with $0\leq \Phi(x)\leq 1$. Figure \ref{fig1} shows a plot of $\Phi(x)$ as a function of $x$.\\

The definition of the metric $D_E$ can be extended to the case of mixed states. Given two arbitrary mixed quantum states represented by density matrices $\rho$ and $\sigma$ belonging to $\mathcal{B}(\mathcal{H})_1^+$, the metric $D_E(\rho,\sigma)$ is defined as follows \cite{Lamb2009}:

\begin{equation}
\label{defDEmix}
D_E(\rho,\sigma) \equiv \min_{\ket{\varphi}} \sqrt{H_N\left( \frac{\ket{\psi}\bra{\psi} + \ket{\varphi}\bra{{\varphi}}}{2} \right)}
\end{equation}

In this last expression, $\ket{\psi}$ represents any {\em fixed} purification of $\rho$, and the minimization is taken over all purifications $\ket{\varphi}$ of $\sigma$.\\
In order to derive some appealing properties of $D_E$ it is useful to write it down in terms of the Uhlmann-Jozsa fidelity $F$ (see appendix):

\begin{equation}
\label{UJFtext}
F(\rho, \sigma) = \max_{\ket{\varphi}}\left|\braket{\psi}{\varphi}\right|^2
\end{equation}

where $\ket{\psi}$ is any {\em fixed} purification of $\rho$ and maximization is performed over all purifications $\ket{\varphi}$ of $\sigma$. Thus, taking into account equations \ref{defDEmix}, \ref{DEPhix} and \ref{UJFtext}, it is straight forward to see that $D_E$ can be expressed as

\begin{equation}
\label{DEvsF}
D_E(\rho,\sigma) = \sqrt{\Phi\left(\sqrt{F(\rho,\sigma)}\right)}
\end{equation}

\subsection{\label{DEprop}Properties of the distance $D_E$}

To easily see that $D_E$ is a metric we can write $D_E$ as a function of the Bures distance $D_B$ taking into account that both distances can be expressed in terms of the fidelity $F$ [cf. Eqs. \ref{DBures} (see appendix) and \ref{DEvsF}]. Thus, we have

\begin{equation}
D_E(D_B) = \sqrt{\Phi\left(1-\frac{D_B^2}{2}\right)}
\end{equation}

where $\Phi(.)$ is given by Eq. \ref{Phix}.

\begin{figure}[h!]
\includegraphics[scale=0.25]{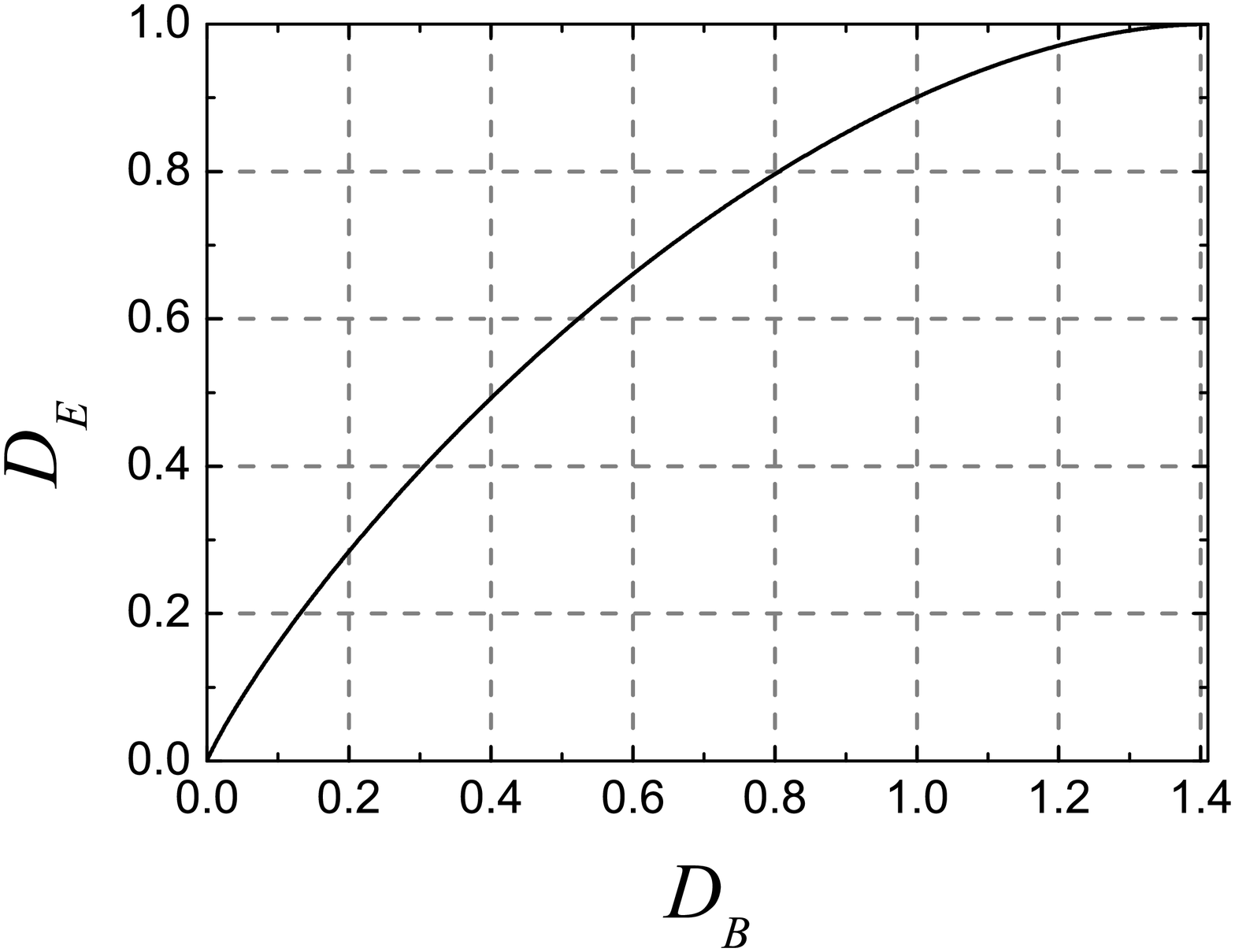}
\caption{\label{fig2}Distance $D_E$ as a function of Bures distance $D_B$}
\end{figure}

Figure \ref{fig2} shows a plot of $D_E$ as a function of $D_B$. As this function is concave, $D_E$ satisfies the properties of a metric.

From its definition [cf. Eqs. \ref{defDEpure}, \ref{defDEmix} and \ref{DEvsF}], it can be formally proved that $D_E$ satisfies the following properties:

\begin{enumerate}

\item \label{PDE1} {\em Normalization}:
\begin{equation}
0\leq D_E(\rho,\sigma)\leq 1
\end{equation}

\item \label{PDE2} {\em Identity of indiscernibles}
\begin{equation}
D_E(\rho, \sigma) = 0 \:\:{\rm if\:\:and\:\:only\:\:if}\:\:\rho = \sigma
\end{equation}
For {\em pures states} $D_E$ vanishes if and only if $\ket{\psi} = e^{\i a} \ket{\varphi}$  (i.e., the two states belong to the same ray in the Hilbert space).\\

\item \label{PDE3} {\em Symmetry}:
\begin{equation}
D_E(\rho,\sigma) = D_E(\sigma,\rho)
\end{equation}

\item \label{PDE4} {\em Triangle inequality}: For any arbitrary density matrices $\rho$, $\sigma$ and $\xi$
\begin{equation}
D_E(\rho,\sigma) \leq D_E(\rho,\xi) + D_E(\xi,\sigma)
\end{equation}

For proofs of properties \ref{PDE1} to \ref{PDE4} refer to Refs. \cite{Lamb2009}, \cite{Lamb2008,Briet} and \cite{Majtey05}.\\

\item \label{PDE5}{\em Joint convexity}: For $p_i \geq 0$, $\sum_i p_i =1$, $\rho_i$ and $\sigma_i$ arbitrary density matrices
\begin{eqnarray}
&& D_E^2(\sum_i p_i \rho_i, \sum_i p_i \sigma_i)\\
&& \leq \sum_i p_i D_E^2(\rho_i,\sigma_i)
\end{eqnarray}

{\em Proof:} It follows from Eq. \ref{DEvsQJSD} and the {\em Joint concavity} property of $D_{JS}$ [cf. Eq. \ref{QJSD}] \cite{Majtey05}.\\
{\em Remark:} Note that {\em joint convexity} implies {\em separate convexity}, but not the converse. For example, the {\em separate convexity} of $D_E^2$ can be obtained from {\em joint convexity} by setting $\sigma_i = \sigma$ and using the fact that $\sum_i p_i=1$.\\

\item \label{PDE6}{\em Restricted additivity}: For any arbitrary density matrices $\rho_1$, $\sigma_1$ and $\tau$ 
\begin{equation}
D_{E}(\rho_1 \otimes \tau, \sigma_1 \otimes \tau) = D_{E}(\rho_1, \sigma_1)
\end{equation}

{\em Proof:} To prove this property we observe Eq. \ref{DEvsF} which relates $D_E$ with the Uhlmann-Jozsa fidelity $F$. Then, we use properties \ref{FP6} and \ref{FP2} of $F$ (cf. appendix). By setting $\rho_2 = \sigma_2 = \tau$ in property \ref{FP6} of $F$ it follows that

\begin{equation}
F(\rho_1 \otimes \tau, \sigma_1 \otimes \tau) = F(\rho_1,\sigma_1)
\end{equation}
This fact completes de proof of the {\em Restricted additivity} property of $D_E$.\\
{\em Comment:} An immediate consequence of this property is that for two physical systems, described by density matrices $\rho_1$ and $\sigma_1$, a measure of their degree of similarity determined by means of $D_E$ remains unchanged even after appending to each system an uncorrelated ancillary state $\tau$.

\item \label{PDE7} {\em Unitary invariance}: For any unitary operation $\mathcal{U}$
\begin{equation}
D_E(\mathcal{U} \rho\, \mathcal{U}^\dag, \mathcal{U} \sigma\, \mathcal{U}^\dag) = D_E(\rho,\sigma)
\end{equation}

{\em Proof:} It follows from Eq. \ref{DEvsF} and the {\em Unitary invariance} property of $F$ (cf. appendix).\\
{\em Comment:} This is a quite natural property to be satisfied by a distance, because a unitary transformation represents a rotation in the Hilbert space and the distance between two states should be invariant under a rotation of the states.

\item \label{PDE8}{\em Monotonicity under quantum operations}: If $\mathcal{E}$ is a CPTP-map, then for any arbitrary density matrices $\rho$ and $\sigma$
\begin{equation}
\label{Hteo}
D_E(\mathcal{E}(\rho),\mathcal{E}(\sigma)) \leq D_E(\rho,\sigma)
\end{equation}

{\em Proof:} It follows from Eq. \ref{DEvsF} and the {\em Monotonicity under quantum operations} property of $F$ (cf. appendix) taking into account that the function $\Phi(.)$ is a monotonically decreasing function of $\sqrt{F}$ [cf. Eq. \ref{Phix}].\\
{\em Comment:} This is a very important property because it qualifies $D_E$ as a monotonically decreasing measure under CPTP maps and can be considered the quantum analog of the classical information-processing inequality which states that the amount of information should not increase via any information processing. For example, the dynamics of an open quantum system can be described by means of a CPTP-map using an operator-sum representation of the form of Eq. \ref{osr}. Therefore the meaning of Eq. \ref{Hteo} is that nonunitary evolution decreases distinguishability between states. Of course, a unitary evolution is a particular case of a CPTP-map. In this case, equality is satisfied in Eq. \ref{Hteo} in complete agreement with property \ref{PDE7}. Another example of a CPTP-map is given by 

\begin{equation}
\mathcal{E}(\rho) = \sum_i P_i \rho P_i
\end{equation}

with $P_i$ being a complete set of orthogonal projectors (i.e., $P_i^\dag = P_i$, $P_i^2 = P_i$ and $\sum_i P_i = I$). In this case, property \ref{PDE8} directly implies

\begin{eqnarray}
D_E^2(\sum_i P_i \rho P_i, \sum_i P_i \sigma P_i)&&\\
&& = \sum_i D_E^2(P_i \rho P_i, P_i \sigma P_i)\\
&& \leq D_E^2(\rho,\sigma)
\end{eqnarray}

Therefore, due to the monotonic character of the square root we have

\begin{equation}
D_E(\sum_i P_i \rho P_i, \sum_i P_i \sigma P_i) \leq D_E(\rho,\sigma)
\end{equation}
\end{enumerate}

\subsection{Physical interpretation of $D_E$}

Quantum Jensen-Shannon divergence $D_{JS}$ [cf. Eq.\ref{QJSD}] can be generalized as a ``measure of distance" between the elements of an ensemble $\lbrace q_i,\rho_i \rbrace$ $(\sum_i q_i = 1)$ \cite{Majtey05}

\begin{equation}
D_{JS}(\lbrace q_i,\rho_i\rbrace) = H_N (\sum_i q_i \rho_i) - \sum_i q_i H_N(\rho_i)
\end{equation}

In the context of quantum transmission processes this quantity represents the Holevo quantity, which bounds the mutual information between the sender of a classical message encoded in quantum states and a receiver.\\
In a recent paper \cite{Roga}, \.Zyczkowski and co-workers showed that the square of the distance $D_E$ provides a finest bound for the Holevo quantity for a particular ensemble $\lbrace q_1 = 1/2, q_2 = 1/2, \rho_1,\rho_2\rbrace$. In this way, the distance $D_E$ turns out to be endowed with an important physical meaning.\\
Another point to be remarked about $D_E$ is related to the fact that this distance could be implemented operationally. Indeed, Ricci et al. \cite{Ricci} reported an experimental implementation of a theoretical protocol for the purification of single qubits sent through a depolarizing channel previously proposed by Cirac and co-workers \cite{Cirac}.

\section{\label{ALTFE}Alternative fidelity definition}

A very interesting and neat feature of the metric $D_E$ is that a fidelity $F_E$ for both pure and mixed quantum states can be defined which fulfills the most important properties satisfied by the usual (Uhlmann-Jozsa) fidelity $F$.
Bearing in mind Ref. \cite{Majtey05}, we define an alternative fidelity measure $F_E$ as follows:

\begin{equation}
\label{fide}
F_E(\rho,\sigma) \equiv \left[ 1 - D_E^2(\rho,\sigma)\right]
\end{equation}

The most important properties of $F_E$ are the following:

\begin{enumerate}
\item \label{PFE1}{\em Normalization}:
\begin{equation}
0 \leq F_E(\rho,\sigma) \leq 1
\end{equation}
$F_E(\rho,\sigma) = 0$ if $\rho$ and $\sigma$ have supports on orthogonal subspaces
\\
{\em Proof:} It follows straight forward from definition of $F_E$ [cf. Eq. \ref{fide}] and the {\em Normalization} property of $D_E$ (cf. Sec. \ref{DEprop}).
\item \label{PFE2} {\em Identity of indiscernibles}:
\begin{equation}
F_E(\rho,\sigma) = 1\:\: {\rm\:\: if\:\: and \:\:only\:\: if}\:\: \rho = \sigma
\end{equation}
\\
{\em Proof:} It follows straight forward from definition of $F_E$ [cf. Eq. \ref{fide}] and the {\em Identity of indiscernibles} property of $D_E$ (cf. Sec. \ref{DEprop}).
\item \label{PFE3}{\em Symmetry}:
\begin{equation}
F_E(\rho,\sigma) = F_E(\sigma, \rho)
\end{equation}
\\
{\em Proof:} It follows straight forward from definition of $F_E$ [cf. Eq. \ref{fide}] and the {\em Symmetry} property of $D_E$ (cf. Sec. \ref{DEprop}).
\item \label{PFE4}{\em Joint Concavity}: For $p_i\geq 0$, $\sum_i p_i=1$, $\rho_i$ and $\sigma_i$ arbitrary density matrices
\begin{eqnarray}
&&F_E(\sum_i p_i \rho_i, \sum_i p_i \sigma_i)\\
&& \geq \sum_i p_i F_E(\rho_i,\sigma_i)
\end{eqnarray}
\\
{\em Proof:} It follows immediately from definition of $F_E$ [cf. Eq. \ref{fide}] and the property of {\em Joint Convexity} of $D_E^2$ (cf. Sec. \ref{DEprop}).

{\em Remark:} While Uhlmann-Jozsa fidelity $F$ has the property of being {\em separate concave} in each of its arguments, $F_E$ turns out to have the enhanced {\em Joint Concavity} property. Therefore, separate concavity on each of its arguments is also satisfied.

\item \label{PFE5}{\em Restricted additivity}:
\begin{equation}
F_E(\rho \otimes \tau, \sigma \otimes \tau) = F_E(\rho,\sigma)
\end{equation}
\\
{\em Proof:} It follows straight forward from definition of $F_E$ [cf. Eq. \ref{fide}] and the {\em Restricted additivity} property of $D_E$ (cf. Sec. \ref{DEprop}).\\
{\em Comment:} As a consequence of this property, a measure of the degree of similarity between two physical systems described by density matrices $\rho$ and $\sigma$ by means of $F_E$ remains unchanged even after appending to each system an uncorrelated ancillary state $\tau$.

\item \label{PFE6} {\em Unitary invariance}: For any unitary operation $\mathcal{U}$
\begin{equation}
F_E(\mathcal{U} \rho\, \mathcal{U}^\dag, \mathcal{U} \sigma\, \mathcal{U}^\dag) = F_E(\rho,\sigma)
\end{equation}

{\em Proof:} It follows straight forward from definition of $F_E$ [cf. Eq. \ref{fide}] and the {\em Unitary invariance} property of $D_E$ (cf. Sec. \ref{DEprop}).\\

\item \label{PFE7}{\em Monotonicity under quantum operations}:
\begin{equation}
F_E(\mathcal{E}(\rho), \mathcal{E}(\sigma)) \geq F_E(\rho,\sigma)
\end{equation}
where $\mathcal{E}$ is a CPTP-map.
\\
{\em Proof:} It follows from definition of $F_E$ [cf. Eq. \ref{fide}] as a direct consequence of the property of {\em Monotonicity under quantum operations} of $D_E$ (cf. section \ref{DEprop}).\\

{\em Comment:} Physically, this property means that, as $F_E$ serves as a kind of measure for the degree of similarity between two quantum states $\rho$ and $\sigma$, one might expect that a general quantum operation $\mathcal{E}$ will make them less distinguishable and, therefore, more similar according to $F_E$. Thus, this property qualifies $F_E$ as a monotonically increasing measure under CPTP-maps. 
\end{enumerate}

\section{\label{DeltaE}Metric to measure distances between quantum processes based on the metric $D_E$}

From the distance $D_E$ it is possible to introduce a distance $\Delta_E$ between quantum processes. Following Gilchrist, Langford and Nielsen \cite{Gilch}, we define the distance $\Delta_E$ between the quantum processes $\mathcal{E}$ and $\mathcal{F}$ as:

\begin{equation}
\label{DeltaN}
\Delta_E(\mathcal{E},\mathcal{F}) \equiv D_E(\rho_{\mathcal{E}}, \rho_{\mathcal{F}})
\end{equation}

where $\rho_{\mathcal{E}}$ and $\rho_{\mathcal{F}}$ are the Jamio{\l}kowski isomorphisms corresponding to the quantum processes $\mathcal{E}$ and $\mathcal{F}$ [cf. Eq. \ref{jamiso}].\\

The fundamental properties of $\Delta_E$ are presented below. It is easy to see that the properties of {\em Normalization} and {\em Symmetry} of $\Delta_E$ are inherited from the corresponding properties of $D_E$.

\begin{enumerate}
\item \label{pp1} {\em Normalization}:
\begin{equation}
0 \leq \Delta_E(\mathcal{E},\mathcal{F}) \leq 1
\end{equation}

\item \label{pp} {\em Identity of indiscernibles}:
\begin{equation}
\Delta_E(\mathcal{E},\mathcal{F}) = 0\:\: {\rm if\:\: and\:\: only\:\: if\:\:} \mathcal{E} = \mathcal{F}
\end{equation}
\\
{\em Proof:} It can be proved recalling the Jamio{\l}kowski isomorphism (cf. Sec. \ref {JAMISO}) and definition of $\Delta_E(\mathcal{E},\mathcal{F})$  [cf. Eq. \ref{DeltaN}]. Thus, there exist an univocal relationship between a quantum process $\mathcal{E}$ and the Jamio{\l}kowski state $\rho_{\mathcal{E}}$. Therefore, if $\mathcal{E} \neq \mathcal{F}$ it follows that $\rho_{\mathcal{E}} \neq \rho_{\mathcal{F}}$ and $\Delta_E$ satisfies property \ref{pp}.\\
\item \label{pp2} {\em Symmetry}:
\begin{equation}
\Delta_E(\mathcal{E},\mathcal{F}) = \Delta_E(\mathcal{F},\mathcal{E})
\end{equation}

\item \label{pp3} {\em Triangle inequality}: For any three quantum processes $\mathcal{E}$, $\mathcal{F}$ and $\mathcal{G}$
\begin{equation}
\Delta_E(\mathcal{E},\mathcal{G})\leq \Delta_E(\mathcal{E},\mathcal{F})+\Delta_E(\mathcal{F},\mathcal{G})
\end{equation}
\\
{\em Proof:} We start from the metric character of $D_E$ (cf. Sec. \ref{DEprop}). Thus, for given processes $\mathcal{E}$, $\mathcal{F}$ and $\mathcal{G}$ with their corresponding Jamio{\l}kowski states $\rho_{\mathcal{E}}$, $\rho_{\mathcal{F}}$ and $\rho_{\mathcal{G}}$ (cf. Sec. \ref {JAMISO}), we have:

\begin{eqnarray}
\Delta_E(\mathcal{E},\mathcal{F})+\Delta_E(\mathcal{F},\mathcal{G}) - \Delta_E(\mathcal{E},\mathcal{G}) &=&\\
D_E(\rho_{\mathcal{E}},\rho_{\mathcal{F}})+D_E(\rho_{\mathcal{F}},\rho_{\mathcal{G}}) - D_E(\rho_{\mathcal{E}},\rho_{\mathcal{G}}) \geq 0
\end{eqnarray}

\item \label{pp4} {\em Stability}:
\begin{equation}
\Delta_E(\mathcal{I} \otimes \mathcal{E}, \mathcal{I} \otimes \mathcal{F}) = \Delta_E(\mathcal{E},\mathcal{F})
\end{equation}
where $\mathcal{I}$ represents the identity operation on an extra Hilbert of arbitrary dimension.
\\
{\em Proof:} We start from definition of $\Delta_E(\mathcal{E},\mathcal{F})$ [cf. Eq. \ref{DeltaN}] and use the property of {\em restricted additivity} of $D_E$ (cf. Sec. \ref{DEprop}). Thus, we have:

\begin{eqnarray}
\Delta_E(\mathcal{I} \otimes \mathcal{E}, \mathcal{I} \otimes \mathcal{F}) & = & D_E(\rho_{\mathcal{I} \otimes \mathcal{E}}, \rho_{\mathcal{I} \otimes \mathcal{F}})\\
& = & D_E(\rho_{\mathcal{I}} \otimes \rho_{\mathcal{E}}, \rho_{\mathcal{I}} \otimes \rho_{\mathcal{F}})\\
& = & D_E(\rho_{\mathcal{E}},\rho_{\mathcal{F}}) = \Delta_E(\mathcal{E},\mathcal{F})
\end{eqnarray}

In last equation we used the useful property $\rho_{\mathcal{E}\otimes \mathcal{F}} = \rho_{\mathcal{E}}\otimes \rho_{\mathcal{F}}$ \cite{Gilch}.\\
\item \label{pp5} {\em Chaining}: For any quantum processes $\mathcal{E}_1$, $\mathcal{E}_2$, $\mathcal{F}_1$ and $\mathcal{F}_2$
\begin{equation}
\Delta_E(\mathcal{E}_2 \circ \mathcal{E}_1, \mathcal{F}_2 \circ \mathcal{F}_1) \leq  \Delta_E(\mathcal{E}_1,\mathcal{F}_1)+\Delta(\mathcal{E}_2,\mathcal{F}_2)
\end{equation}
\end{enumerate}
{\em Proof:} We use the {\em contractivity} property of $D_E$ and, additionally, we assume that $\mathcal{F}_1$ is doubly stochastic, i.e., $\mathcal{F}_1$ is trace-preserving and satisfies $\mathcal{F}_1(I) = I$ (cf. Sec. \ref{OPSR}). This is not a significant assumption, since in quantum information science we are typically interested in the case when $\mathcal{F}_1$ and $\mathcal{F}_2$ are ideal {\em unital} processes, and we want to use $\Delta_E$ to compare the composition of these two ideal processes to the experimentally realized process $\mathcal{E}_2 \circ \mathcal{E}_1$.\\
The proof of the {\em chaining} property starts by applying property \ref{pp3}, i.e., {\em triangle inequality}, so we have:

\begin{eqnarray}
\nonumber
\Delta_E(\mathcal{E}_2 \circ \mathcal{E}_1,\mathcal{F}_2 \circ \mathcal{F}_1) = D_E(\rho_{\mathcal{E}_2 \circ \mathcal{E}_1},\rho_{\mathcal{F}_2 \circ \mathcal{F}_1})&&\\
\label{STRH}
\leq D_E(\rho_{\mathcal{E}_2 \circ \mathcal{E}_1},\rho_{\mathcal{E}_2 \circ \mathcal{F}_1}) + D_E(\rho_{\mathcal{E}_2 \circ \mathcal{F}_1},\rho_{\mathcal{F}_2 \circ \mathcal{F}_1})&&
\end{eqnarray}

Then, we note the easily verified indentity $\rho_{\mathcal{E} \circ \mathcal{F}} = [\mathcal{F}^T \otimes \mathcal{E}] \rho_\Psi$ \cite{Gilch}, where $\rho_\Psi = \ketbra{\Psi}{\Psi}$ with $\ket{\Psi}$ being the the maximally entangled state [cf. Eq. \ref{maxentang}]. Next, we define $\mathcal{F}^T(\rho) = \sum_j F_j^T \rho F_j^*$, where $F_j$ are the operation elements for $\mathcal{F}$ [cf. Eq. \ref{osr}]. Applying this identity to both density matrices in the second term on the right-hand side of Eq. \ref{STRH} we obtain

\begin{eqnarray}
\nonumber
&&\Delta_E(\mathcal{E}_2 \circ \mathcal{E}_1,\mathcal{F}_2 \circ \mathcal{F}_1)\\
\nonumber
&&\leq D_E(\rho_{\mathcal{E}_2 \circ \mathcal{E}_1},\rho_{\mathcal{E}_2 \circ \mathcal{F}_1})\\
\label{FSTRH}
&& + D_E\left([\mathcal{F}_1^T \otimes \mathcal{E}_2]\rho_\Psi,[\mathcal{F}_1^T \otimes \mathcal{F}_2]\rho_\Psi\right)
\end{eqnarray}

The double stochasticity of $\mathcal{F}_1$ implies that $\mathcal{F}_1^T$ is a trace-preserving quantum operation. Then, to complete the proof, we can apply the contractivity property of $D_E$ to both the first and the second terms on the right-hand side of Eq. \ref{FSTRH}.

\begin{eqnarray}
\Delta_E(\mathcal{E}_2 \circ \mathcal{E}_1,\mathcal{F}_2 \circ \mathcal{F}_1) &\leq& D_E(\rho_{\mathcal{E}_1},\rho_{\mathcal{F}_1}) + D_E(\rho_{\mathcal{E}_2},\rho_{\mathcal{F}_2})\\
& = & \Delta_E(\mathcal{E}_1,\mathcal{F}_1)+\Delta(\mathcal{E}_2,\mathcal{F}_2)
\end{eqnarray}

In addition, since unitary processes are also doubly stochastic, it follows that chaining holds for $\Delta_E$ in most cases of usual interest.\\

{\em Remark}: Some interesting properties can be derived from the preceding ones. For example, from the metric and chaining properties it is possible to show that \cite{Gilch}
\begin{equation}
\Delta_E(\mathcal{R}\circ \mathcal{E}, \mathcal{R}\circ \mathcal{F}) \leq \Delta_E(\mathcal{E},\mathcal{F})
\end{equation}

where $\mathcal{R}$ is any quantum operation. Physically, this means that postprocessing $\mathcal{E}$ by $\mathcal{R}$ cannot increase the distinguishability of two processes $\mathcal{E}$ and $\mathcal{F}$. Another interesting consequence of the metric and chaining criteria is the property of {\em unitary invariance}, i.e.,

\begin{equation}
\Delta_E(\mathcal{U} \circ \mathcal{E} \circ \mathcal{V}, \mathcal{U} \circ \mathcal{F} \circ \mathcal{V}) = \Delta_E(\mathcal{E},\mathcal{F})
\end{equation}

where $\mathcal{U}$ and $\mathcal{V}$ are arbitrary unitary operations \cite{Gilch}.

\section{\label{CONCREM}Concluding remarks}

The main results of this paper are concerned with the properties of the entropic metric $D_E$ between quantum states proposed in Ref. \cite{Lamb2009}. Our results indicate that $D_E$ and the derived metric $\Delta_E$ show interesting and useful properties to measure distances between quantum states and quantum processes, respectively. These properties, in general, do not depend on the particular quantum system or process to be considered (as it was emphasized in Sec. \ref{DeltaE}). In addition, we derived an alternative measure of fidelity $F_E$ between quantum states which present the most important properties of the Uhlmann-Jozsa fidelity $F$ such as {\em normalization}, {\em Symmetry} and {\em Monotonicity under quantum operations}. Moreover, the derived fidelity $F_E$ shows the enhanced property of {\em Joint Concavity} with respect to the fidelity $F$ which present the property of {\em Separate Concavity}. 
Regarding practical calculations of the metric $D_E$, in Ref. \cite{Lamb2009} it is shown how to apply this metric to calculate the distance between a mixed qubit and the resulting state when this qubit is sent through a depolarizing channel. Besides, from an experimental viewpoint, it is important to mention that an experimental realization of a theoretical purification protocol \cite{Cirac} has been already achieved in the case of photons sent through a depolarizing channel \cite{Ricci}. These results are very promising because they open a window to think of the possibility of using $D_E$ directly from purifications of quantum states experimentally obtained. Furthermore, it is important to mention that Roga, Fannes and \.Zyczkowski already found a finest bound for the Holevo quantity which turns out to be the square of the $D_E$ metric \cite{Roga}. These facts encourage us to continue investigating how to apply this metric to different cases of interest beyond the depolarizing channel. This task is currently in progress. Certainly, the possibility of evaluating $D_E$ in as many contexts as possible is of central importance. Particular applications to quantum noisy channels represented by sums of operators belonging to Pauli group will be also an interesting matter of study \cite{Niel,Bombin12}. Applications to topological insulators will be also matter of consideration \cite{Hasan,Qi,Viyu12}. Some advances in the context of quantum operations written in the operator-sum representation \cite{Niel} have been made and the results will be presented elsewhere.\\

\section*{Appendix: Uhlmann-Jozsa Fidelity}

In literature, the Uhlmann-Jozsa fidelity $F$ is a celebrated and widely used measure of the degree of similarity between two general density matrices. Fidelity $F$ is given by \cite{Uhlm,Jozsa}

\begin{equation}
\label{UJFdef1}
F(\rho, \sigma) = \left[ \tr \left( \sqrt{\sqrt{\rho} \sigma \sqrt{\rho}} \right) \right]^2
\end{equation}

where $\rho$ and $\sigma$ are arbitrary density matrices.\\
An equivalent definition of $F$ can be provided in terms of purifications of the states $\rho$ and $\sigma$ \cite{Jozsa}

\begin{equation}
\label{UJFdef2}
F(\rho, \sigma) = \max_{\ket{\varphi}}\left|\braket{\psi}{\varphi}\right|^2
\end{equation}

where $\ket{\psi}$ is any {\em fixed} purification of $\rho$ and maximization is performed over all purifications $\ket{\varphi}$ of $\sigma$.\\
For easy access, we summarize below the most appealing properties of Uhlmann-Jozsa fidelity $F$ and adequate properties names and definitions according to the context of the present work. These properties are used in Secs. \ref{DEprop} and \ref{DeltaE} to analyze the properties of the distance $D_E$ and the distance between quantum processes $\Delta_E$ that we will introduce in this work. For proofs of the properties listed here see, for example, Refs. \cite{Jozsa,Niel}.\\

\begin{enumerate}
\item \label{FP1} {\em Normalization:}
\begin{equation}
0 \leq F(\rho, \sigma) \leq 1
\end{equation}
\item \label{FP2} {\em Identity of indiscernibles:}
\begin{equation}
F(\rho, \sigma) = 1 \:\:{\rm if\:\:and\:\:only\:\:if}\:\:\rho = \sigma
\end{equation}
\item \label{FP3} {\em Symmetry:}
\begin{equation}
F(\rho, \sigma) = F(\sigma, \rho)
\end{equation}
\item \label{FP4} If $\rho = \ketbra{\xi}{\xi}$ represents a pure state then $F(\rho, \sigma) = \bra{\xi}\sigma\ket{\xi} = \tr{(\rho \,\sigma)}$
\item \label{FP5}{\em Separate Concavity:} For $p_1 , p_2 \geq 0$, $p_1+p_2=1$, and arbitrary density matrices $\rho_1$, $\rho_2$ and $\sigma$
\begin{equation}
F(p_1 \rho_1+p_2 \rho_2, \sigma) \geq p_1 F(\rho_1,\sigma) +p_2 F(\rho_2,\sigma)
\end{equation}
By symmetry property \ref{FP3}, concavity in the second argument is also fulfilled.
\item \label{FP6} {\em Multiplicativity under tensor product:} For arbitrary density matrices $\rho_1$, $\rho_2$, $\sigma_1$ and $\sigma_2$
\begin{equation}
F(\rho_1 \otimes \rho_2, \sigma_1 \otimes \sigma_2) = F(\rho_1,\sigma_1) F(\rho_2, \sigma_2)
\end{equation}
\item \label{FP7} {\em Unitary invariance:} For any arbitrary unitary process $\mathcal{U}$, $F(\rho, \sigma)$ is preserved i.e., 
\begin{equation}
F(\mathcal{U} \rho\, \mathcal{U}^\dag), \mathcal{U} \sigma \mathcal{U}^\dag) = F(\rho, \sigma)
\end{equation}
\item \label{FP8} {\em Monotonicity under quantum operations:} For a general quantum operation $\mathcal{E}$ described by a CPTP-map (cf. section \ref{OPSR})
\begin{equation}
F(\mathcal{E}(\rho), \mathcal{E}(\sigma)) \geq F(\rho, \sigma)
\end{equation}
\end{enumerate}

{\em Remark 1}: The fidelity $F$ serves as a generalized measure of the overlap between two quantum states but is not a metric. However, the fidelity can easily be turned into a metric. For example, the Bures distance is a {\em metric} which can be defined in terms of the fidelity $F$ as \cite{Bengt}:

\begin{equation}
\label{DBures}
D_B(\rho,\sigma) = \sqrt{2 - 2 \sqrt{F(\rho, \sigma)}}
\end{equation}

{\em Remark 2}: While $F$ satisfies {\em separate concavity} it can be shown that $\sqrt{F}$ is {\em jointly concave} \cite{Uhlm00,Mendo} i.e.,
\begin{eqnarray}
\nonumber
&&\sqrt{F(p_1 \rho_1+p_2 \rho_2, p_1 \sigma_1 + p_2 \sigma_2)}\\
\label{jointconc}
&& \geq p_1 \sqrt{F(\rho_1,\sigma_1)} + p_2 \sqrt{F(\rho_2,\sigma_2)}
\end{eqnarray}
where $p_1 , p_2 \geq 0$, $p_1+p_2=1$, and $\rho_1$, $\rho_2$, $\sigma_1$ and $\sigma_2$ are arbitrary density matrices.

{\em Remark 3}: Clearly, by extension, $\sqrt{F}$ satisfies all properties of the fidelity $F$ but property \ref{FP4}.

{\em Remark 4}:  It is important to realize that any measure $M$ which is {\em unitarily invariant}, {\em jointly concave (convex)}, and {\em invariant under the addition of an ancillary system} is also monotonically increasing (decreasing) under CPTP-maps \cite{Mendo}, therefore, it turns out to be a suitable measure of the degree of similarity between quantum states.\\

\acknowledgments
T.M.O and P.W.L are fellows of the National Research Council of Argentina (CONICET). The authors are grateful to Secretaria de Ciencia y T\'ecnica de la Universidad Nacional de C\'ordoba (SECyT-UNC, Argentina) for financial support.


\begin{thebibliography}{10}
\bibitem{postsel} Post-selection is usually associated with obtainig some specific outcome after a measurement procedure is performed.
\bibitem{notejam} An important remark, however, is that not all density matrices may be realized as states $\rho_{\mathcal{E}}$. It turns out that the class $\mathcal{E}$ of quantum states that may arise in this way from a trace-preserving operation is simply all those $\rho_{\mathcal{E}}$ whose reduced density matrix on the copy of the original system is the completely mixed state, $I/d$ \cite{Horo}. This fact does not constitutes a drawback for the purpose of this paper since we are interested on mapping a quantum process $\mathcal{E}$ into a density matrix $\rho_{\mathcal{E}}$ and not the converse.
\end{thebibliography}
\end{document}